# Duffin-Kemmer-Petiau particle in a vector exponential-like decaying field with any arbitrary $J$-state

S. M. Ikhdair

*Physics Department, Near East University, 922022 Nicosia, North Cyprus, Mersin 10, Turkey*

*and*

*Physics Department, Faculty of Science, An-Najah National University, Nablus, West Bank, Palestine*

Email: sikhdair@neu.edu.tr

**Abstract**

The Duffin–Kemmer–Petiau (DKP) equation is solved approximately for a vector exponential-like decaying potential with any arbitrary $J$-state by using the Pekeris approximation. The generalized parametric Nikiforov-Uvarov (NU) method is used to obtain energy eigenvalues and corresponding wave functions in a closed form. The cases of zero total angular momentum and nonrelativistic limit are discussed too.

## 1. Introduction

The first-order Duffin–Kemmer–Petiau (DKP) formalism which describes spin-0 and spin-1 particles has been used to analyze relativistic interactions of spin-0 and spin-1 hadrons with nuclei as an alternative to their conventional second-order Klein-Gordon and Proca counterparts [1-5]. The DKP equation is a direct generalization to the Dirac particles of integer spin in which one replaces the gamma matrices by beta metrics but verifying a more complicated algebra as DKP algebra [6-14]. Fainberg and Pimentel presented a strict proof of equivalence between DKP and Klein-Gordon theories for physical S-matrix elements in the case of charged scalar particles interacting in minimal way with an external or quantized electromagnetic field [15-16]. Boutabia-Chéraitia and Boudjedaa solved the DKP equation in the presence of Woods–Saxon potential for spin 1 and spin 0 and they also deduced the transmission and reflection coefficients [17]. Kulikov et al. offered a new oscillator model with different form of the non-minimal substitution within the framework of the DKP equation [18]. Yaşuk et al. presented an application of the relativistic DKP equation in the presence of a vector deformed Hulthén potential for spin zero particles by using the Nikiforov-Uvarov (NU) method [19]. Boztosun et al. presented a simple exact analytical solution of the relativistic DKP equation within the framework of the asymptotic iteration method and determined exact bound state energy eigenvalues and corresponding wave functions for the relativistic harmonic oscillator as well as the Coulomb potentials [20]. Kasri and Chetouani determined the bound state energy eigenvalues for the relativistic DKP oscillator and DKP Coulomb potentials by using an exact quantization rule [21]. de Castro explored the problem of spin-0 and spin-1 bosons subject to a general mixing of minimal and non-minimal vector cusp potentials in a unified way in the context of the DKP theory [22]. Chagui et al. solved the DKP equation with a pseudo-scalar linear plus Coulomb-like potential in a two-dimensional space–time [23]. For more review, one can see Refs. [24-28].

Very recently, using a Pekeris approximation, the approximate solution of the DKP equation for the vector deformed Woods-Saxon potential [29] and the vector Yukawa potential [30] have been solved with arbitrary total angular momentum $J \neq 0$ in the framework of the parametric generalization of the NU method.

The aim of this paper is to solve the DKP equation for a vector exponential-like decaying potential when the total angular momentum is non-zero, i.e. $J \neq 0$. Under these conditions, the DKP equation cannot be solved exactly because of the angular momentum term, $J(J+1)/r^2$, and hence we must use an approximation scheme to deal with this term.

This paper is organized as follows. In section 2, the DKP formalism is briefly given and discussed under a vector potential. In section 3, parametric generalization of the NU method is introduced. In section 4, using the Pekeris approximation, we obtain an approximate solution of the DKP equation for a vector exponential-like decaying potential with any arbitrary $J$-state via the NU method. Finally, in section 5 we give our concluding remarks.

**2. Review to the DKP Formalism**

The first order relativistic DKP equation for a free spin-0 or spin-1 particle of mass $m$ is

$$(i\beta^\mu \partial_\mu - m)\psi_{DKP} = 0, \tag{1}$$

where $\beta^\mu$ ($\mu = 0,1,2,3$) matrices satisfy the commutation relation

$$\beta^\mu \beta^\nu \beta^\lambda + \beta^\lambda \beta^\nu \beta^\mu = g^{\mu\nu}\beta^\lambda + g^{\nu\lambda}\beta^\mu, \tag{2}$$

which defines the so-called DKP algebra. The algebra generated by the $4\beta^N$'s has three irreducible representations: a ten dimensional one that is related to $S = 1$, a five dimensional one relevant for $S = 0$ (spinless particles) and a one dimensional one which is trivial. In the spin-0 representation, $\beta^\mu$ are $5 \times 5$ matrices defined as ($i = 1,2,3$)

$$\beta^0 = \begin{pmatrix} \theta & \bar{0} \\ \bar{0}_T & 0 \end{pmatrix}, \quad \beta^i = \begin{pmatrix} 0 & \rho^i \\ -\rho^i_T & 0 \end{pmatrix}, \tag{3}$$

with $\tilde{0}$, $\bar{0}$, $0$ as $2 \times 2$, $2 \times 3$, $3 \times 3$ zero matrices, respectively, and

$$\theta = \begin{pmatrix} 0 & 1 \\ 1 & 0 \end{pmatrix}, \quad \rho^1 = \begin{pmatrix} -1 & 0 & 0 \\ 0 & 0 & 0 \end{pmatrix}, \quad \rho^2 = \begin{pmatrix} 0 & -1 & 0 \\ 0 & 0 & 0 \end{pmatrix}, \quad \rho^3 = \begin{pmatrix} 0 & 0 & -1 \\ 0 & 0 & 0 \end{pmatrix}. \tag{4}$$

For spin one particle, $\beta^\mu$ are $10 \times 10$ matrices given by

$$\beta^0 = \begin{pmatrix} 0 & \bar{0} & \bar{0} & \bar{0} \\ \bar{0}^T & 0 & I & 0 \\ \bar{0}^T & I & 0 & 0 \\ \bar{0}^T & 0 & 0 & 0 \end{pmatrix}, \quad \beta^i = \begin{pmatrix} 0 & \bar{0} & e_i & \bar{0} \\ \bar{0}^T & 0 & 0 & -is_i \\ -e_i^T & 0 & 0 & 0 \\ \bar{0}^T & -is_i & 0 & 0 \end{pmatrix}, \tag{5}$$

where $s_i$ are the usual $3 \times 3$ spin one matrices, $I$ and $0$ are the identity and zero matrices, respectively. Further, we have

$$\bar{0} = \begin{pmatrix} 0 & 0 & 0 \end{pmatrix}, \quad e_1 = \begin{pmatrix} 1 & 0 & 0 \end{pmatrix}, \quad e_2 = \begin{pmatrix} 0 & 1 & 0 \end{pmatrix}, \quad e_3 = \begin{pmatrix} 0 & 0 & 1 \end{pmatrix}. \tag{6}$$

While the dynamical state $\psi_{DKP}$ is a five component spinor for spin zero particles, it has ten component spinors for $S = 1$ particles. The solution of the DKP equation for a particle in a central field needs consideration since earlier work [11]. It is convenient to recall some general properties of the solution of the DKP equation in a central interaction for spin zero particles. The central interaction consists of two parts: $U_S(\vec{r})$ stands for a Lorentz scalar and $U_V(\vec{r})$ stands for the time component of the 4-vector potential with a vanishing space component. The stationary states of the DKP particle (in units $\hbar = c = 1$) in this case are determined by solving the following equation:

$$\left(\vec{\beta}.\vec{p} + m + U_S(\vec{r}) + \beta^0 U_V(\vec{r})\right)\psi(\vec{r}) = \beta^0 E \psi(\vec{r}). \tag{7}$$

In the spin zero representation, the five component DKP spinor

$$\psi(\vec{r}) = \begin{pmatrix} \psi_{upper} \\ i\psi_{lower} \end{pmatrix} \text{ with } \psi_{upper} = \begin{pmatrix} \phi \\ \varphi \end{pmatrix} \text{ and } \psi_{lower} = \begin{pmatrix} A_1 \\ A_2 \\ A_3 \end{pmatrix}, \tag{8}$$

so that for stationary states the DKP equation can be written as

$$(m + U_S)\phi = (E - U_V^0)\varphi + \vec{\nabla}.\vec{A}, \tag{9}$$

$$\vec{\nabla}\phi = (m + U_S)\vec{A}, \tag{10}$$

$$(m + U_S)\varphi = (E - U_V^0)\phi, \tag{11}$$

where $\vec{A}$ is the vector $(A_1, A_2, A_3)$. The five-component wave function $\psi$ is simultaneously an eigenfunction of $J^2$ and $J_3$

$$J^2 \begin{pmatrix} \psi_{upper} \\ \psi_{lower} \end{pmatrix} = \begin{pmatrix} L^2 \psi_{upper} \\ (\vec{L} + \vec{S})^2 \psi_{lower} \end{pmatrix} = J(J+1) \begin{pmatrix} \psi_{upper} \\ \psi_{lower} \end{pmatrix}, \tag{12}$$

$$J_3 \begin{pmatrix} \psi_{upper} \\ \psi_{lower} \end{pmatrix} = \begin{pmatrix} L_3 \psi_{upper} \\ (L_3 + S_3) \psi_{lower} \end{pmatrix} = M \begin{pmatrix} \psi_{upper} \\ \psi_{lower} \end{pmatrix}, \tag{13}$$

where the total angular momentum $J = L + S$ which commutes with $\beta^0$, is a constant of the motion. The most general solution of Eq. (7) is

$$\psi_{JM}(r) = \begin{pmatrix} f_{nJ}(r) Y_{JM}(\Omega) \\ g_{nJ}(r) Y_{JM}(\Omega) \\ i \sum_L h_{nJL}(r) Y_{JL1}^M(\Omega) \end{pmatrix} \tag{14}$$

where $Y_{JM}(\Omega)$ are the spherical harmonics of order $J$, $Y_{JL1}^M(\Omega)$ are the normalized vector spherical harmonics and $f_{nJ}(r)$, $g_{nJ}(r)$ and $h_{nJL}(r)$ are radial wave functions. Now, inserting $\psi_{JM}(r)$ given by Eq. (14) into Eqs. (9) - (11) by using the properties of vector spherical harmonics [2] one gets the following set of first-order coupled relativistic differential radial equations

$$(E - U_V^0)F(r) = (m + U_S)G(r), \tag{15a}$$

$$\left(\frac{d}{dr}-\frac{J+1}{r}\right)F(r)=-\frac{1}{\alpha_J}(m+U_S)H_1(r), \tag{15b}$$

$$\left(\frac{d}{dr}+\frac{J}{r}\right)F(r)=-\frac{1}{\varsigma_J}(m+U_S)H_{-1}(r), \tag{15c}$$

$$-\alpha_J\left(\frac{d}{dr}+\frac{J+1}{r}\right)H_1(r)+\varsigma\left(\frac{d}{dr}-\frac{J}{r}\right)H_{-1}(r) \tag{15d}$$
$$=(m+U_S)F(r)-(E-U_V^0)G(r),$$

where $\alpha_J=\sqrt{(J+1)/(2J+1)}$, $\varsigma=\sqrt{J/(2J+1)}$, $f_{nJ}(r)=F(r)/r$, $g_{nJ}(r)=G(r)/r$ and $h_{nJJ\pm1}(r)=H_{\pm1}(r)/r$. Note that the five radial components in the wave function (14) end up with four components $\{F,G,H_{+1},H_{-1}\}$ since the fifth component $H_0$ disappears due to the property $J^2H_J=J(J+1)H_J$, i.e., $J^2H_0=0H_0$ and the multiplication process of Eq. (5) with the wave function (14) after using Eq. (7) (cf. Ref. [25]). For DKP equation, at the presence of vector potential and while scalar potential is zero, the differential equations to be satisfied by the radial wave functions are

$$(E-U_V^0)F(r)=mG(r), \tag{16a}$$

$$\left(\frac{d}{dr}-\frac{J+1}{r}\right)F(r)=-\frac{1}{\alpha_J}mH_1(r), \tag{16b}$$

$$\left(\frac{d}{dr}+\frac{J}{r}\right)F(r)=-\frac{1}{\varsigma_J}mH_{-1}(r), \tag{16c}$$

$$-\alpha_J\left(\frac{d}{dr}+\frac{J+1}{r}\right)H_1(r)+\varsigma\left(\frac{d}{dr}-\frac{J}{r}\right)H_{-1}(r) \tag{16d}$$
$$=mF(r)-(E-U_V^0)G(r).$$

Eliminating $G(r)$, $H_1(r)$ and $H_{-1}(r)$ in terms of $F(r)$ the following second-order equation is obtained

$$\left[\frac{d^2}{dr^2}-\frac{J(J+1)}{r^2}+\left(E-U_V^0\right)^2-m^2\right]F(r)=0, \tag{17}$$

which appears as the radial KG equation for a vector potential.

At this stage, we take the vector potential, $U_V^0$, in Eq. (17) as an exponential-like decaying potential field given by

$$U_V^0(r)=-De^{-a(r-r_0)}, \tag{18}$$

where $D$, $r_0$ and $a$ are constant coefficients. Substituting Eq. (18) into Eq. (17), the radial DKP equation for $F(r)$ reduces to a more simple form:

$$\left[\frac{d^2}{dr^2} - \frac{J(J+1)}{r^2} + D^2 e^{-2a(r-r_0)} + 2EDe^{-a(r-r_0)} + E^2 - m^2\right] F(r) = 0. \tag{19}$$

**2.3. Pekeris Approximation to the Centrifugal Term**

The DKP equation (19) can not be solved analytically for $J \neq 0$ due to the total angular momentum term $J(J+1)/r^2$. Therefore, we need to use a Pekeris approximation [29-37] in order to deal with this term. In this approximation, the centrifugal potential is expanded around $r = r_0$ in a series of powers of $x = (r - r_0)/r_0$ as

$$V_{SO}(r) = \frac{J(J+1)}{r^2} = \frac{J(J+1)}{r_0^2(1+x)^2} = \frac{J(J+1)}{r_0^2}(1 - 2x + 3x^2 - 4x^3 + \ldots). \tag{20}$$

So it is sufficient to keep expansion terms only up to the second order [32]. The following form of the potential can be used instead of the centrifugal potential in the Pekeris approximation [31]:

$$\tilde{V}_{SO}(r) = \frac{J(J+1)}{r_0^2}\left(c_0 + c_1 e^{-\alpha x} + c_2 e^{-2\alpha x}\right), \tag{21}$$

where $\alpha = a r_0$, $c_i$ is the parameter of coefficients ($i = 0,1,2$) and the expression of Eq. (21) can be expanded up to the terms $x^3$ as

$$\tilde{V}_{SO}(r) = \frac{J(J+1)}{r_0^2}\left[c_0 + c_1\left(1 - \alpha x + \frac{\alpha^2 x^2}{2!} - \frac{\alpha^3 x^3}{3!} + \ldots\right) \right. \\ \left. + c_2\left(1 - 2\alpha x + \frac{4\alpha^2 x^2}{2!} - \frac{8\alpha^3 x^3}{3!} + \ldots\right)\right]. \tag{22}$$

After making some arrangements on Eq. (22), we can finally obtain

$$\tilde{V}_{SO}(r) = \frac{J(J+1)}{r_0^2}\left[c_0 + c_1 + c_2 - (c_1\alpha + 2c_2\alpha)x \right. \\ \left. + \left(\frac{1}{2}c_1\alpha^2 + 2c_2\alpha^2\right)x^2 - \left(\frac{1}{6}c_1\alpha^3 + \frac{4}{3}c_2\alpha^2\right)x^3 + \ldots\right]. \tag{23}$$

Thus, comparing equal powers of Eq. (20) with Eq. (23), we obtain the relations between the coefficients $c_i$ ($i = 0,1,2$) and the parameter $\alpha$ as follows:

$$c_0 = 1 - \frac{3}{\alpha} + \frac{3}{\alpha^2},$$
$$c_1 = \frac{4}{\alpha} - \frac{6}{\alpha^2}, \quad (24)$$
$$c_2 = \frac{1}{\alpha} + \frac{3}{\alpha^2}.$$

Now, we can take the potential $\tilde{V}_{SO}(r)$ given by Eq. (21) instead of the total angular momentum potential (20). Hence, by substituting Eq. (21) into Eq. (19), we obtain

$$\left[ \frac{1}{r_0^2} \frac{d^2}{dx^2} - \frac{J(J+1)}{r_0^2} \left( c_0 + c_1 e^{-\alpha x} + c_2 e^{-2\alpha x} \right) \right.$$
$$\left. + D^2 e^{-2\alpha x} + 2ED e^{-\alpha x} + E^2 - m^2 \right] F(x) = 0. \quad (25)$$

In the following section, we shall use the generalized parametric NU method to find solution for Eq. (25)..

## 3. Nikiforov-Uvarov Method

To solve second order differential equations, the NU method can be used with an appropriate coordinate transformation $s = s(r)$ [38]

$$\psi_n''(s) + \frac{\tilde{\tau}(s)}{\sigma(s)} \psi_n'(s) + \frac{\tilde{\sigma}(s)}{\sigma^2(s)} \psi_n(s) = 0, \quad (26)$$

where $\sigma(s)$ and $\tilde{\sigma}(s)$ are polynomials, at most of second-degree, and $\tilde{\tau}(s)$ is a first-degree polynomial. The following equation is a general form of the Schrödinger-like equation written for any potential [39,40]

$$\left[ \frac{d^2}{ds^2} + \frac{\alpha_1 - \alpha_2 s}{s(1 - \alpha_3 s)} \frac{d}{ds} + \frac{-\xi_1 s^2 + \xi_2 s - \xi_3}{\left[ s(1 - \alpha_3 s) \right]^2} \right] \psi_n(s) = 0. \quad (27)$$

According to the parametric generalization of the NU method, the wave functions and the energy eigenvalue equation, respectively, are found as [39,40]

$$\psi(s) = s^{\alpha_{12}} (1 - \alpha_3 s)^{-\alpha_{12} - \frac{\alpha_{13}}{\alpha_3}} P_n^{(\alpha_{10}-1, \frac{\alpha_{11}}{\alpha_3} - \alpha_{10} - 1)} (1 - 2\alpha_3 s), \quad (28)$$

$$\alpha_2 n - (2n+1)\alpha_5 + (2n+1)\left( \sqrt{\alpha_9} + \alpha_3 \sqrt{\alpha_8} \right) + n(n-1)\alpha_3$$
$$+ \alpha_7 + 2\alpha_3 \alpha_8 + 2\sqrt{\alpha_8 \alpha_9} = 0, \quad (29)$$

where

$$\alpha_4 = \frac{1}{2}(1-\alpha_1), \qquad \alpha_5 = \frac{1}{2}(\alpha_2 - 2\alpha_3),$$
$$\alpha_6 = \alpha_5^2 + \xi_1, \qquad \alpha_7 = 2\alpha_4\alpha_5 - \xi_2,$$
$$\alpha_8 = \alpha_4^2 + \xi_3, \qquad \alpha_9 = \alpha_6 + \alpha_3\alpha_7 + \alpha_3^2\alpha_8, \qquad (30)$$

and

$$\alpha_{10} = \alpha_1 + 2\alpha_4 + 2\sqrt{\alpha_8}, \qquad \alpha_{11} = \alpha_2 - 2\alpha_5 + 2\left(\sqrt{\alpha_9} + \alpha_3\sqrt{\alpha_8}\right),$$
$$\alpha_{12} = \alpha_4 + \sqrt{\alpha_8}, \qquad \alpha_{13} = \alpha_5 - \left(\sqrt{\alpha_9} + \alpha_3\sqrt{\alpha_8}\right). \qquad (31)$$

In some cases $\alpha_3 = 0$ and for this type the composites of the wave functions (28) change as

$$\lim_{\alpha_3 \to 0} P_n^{(\alpha_{10}-1, \frac{\alpha_{11}}{\alpha_3}-\alpha_{10}-1)}(1-\alpha_3 s) = L_n^{\alpha_{10}-1}(\alpha_{11} s), \qquad (32)$$

and

$$\lim_{\alpha_3 \to 0}(1-\alpha_3 s)^{-\alpha_{12}-\frac{\alpha_{13}}{\alpha_3}} = e^{\alpha_{13} s}. \qquad (33)$$

Hence, the solution given by Eq. (28) becomes as [39]

$$\psi(s) = s^{\alpha_{12}} e^{\alpha_{13} s} L_n^{\alpha_{10}-1}(\alpha_{11} s) \qquad (34)$$

## 4. Solution of the DKP Equation for a Vector Exponential-Like Potential

We are now going to solve Eq. (25) for a vector exponential-like decaying potential in the framework of the parametric generalization of the NU method. Thus, by using a convenient change of variables, $s = e^{-\alpha x}$, we rewrite Eq. (25) as follows

$$\frac{d^2 F(s)}{ds^2} + \frac{1}{s}\frac{dF(s)}{ds} + \frac{1}{\alpha^2}\left[-J(J+1)(c_0 + c_1 s + c_2 s^2)\right.$$
$$\left. + 2EDr_0^2 s + D^2 r_0^2 s^2 + (E^2 - m^2)r_0^2\right] F(s) = 0. \qquad (35)$$

Further, comparing Eq. (35) with Eq. (27), we can easily obtain the coefficients $\alpha_i$ ($i = 1, 2, 3$) and analytical expressions for $\xi_j$ ($j = 1, 2, 3$) as follows

$$\alpha_1 = 1, \qquad \xi_1 = \frac{1}{\alpha^2}\left(J(J+1)c_2 + D^2 r_0^2\right),$$
$$\alpha_2 = 0, \qquad \xi_2 = \frac{1}{\alpha^2}\left(-J(J+1)c_1 + 2EDr_0^2\right),$$
$$\alpha_3 = 0, \qquad \xi_3 = \frac{1}{\alpha^2}\left(J(J+1)c_0 - (E^2 - m^2)r_0^2\right). \qquad (36)$$

Overmore, the values of the coefficients $\alpha_i$ ($i = 4, 5, ..., 13$) are also found from the relations (30) and (31). The specific values of the coefficients $\alpha_i$ ($i = 1, 2, ..., 13$) together $\xi_j$ ($j = 1, 2, 3$) are displayed in table 1. By using (29), we can obtain the closed form of the energy eigenvalue equation as

$$(2n+1)\sqrt{\xi_1} - \xi_2 + 2\sqrt{\xi_1 \xi_3} = 0, \tag{37}$$

or, equivalently

$$(2n+1)\alpha\sqrt{(J(J+1)c_2 + D^2 r_0^2)} + (J(J+1)c_1 - 2EDr_0^2)$$
$$+ 2\sqrt{(J(J+1)c_2 + D^2 r_0^2)(J(J+1)c_0 + (m^2 - E^2)r_0^2)} = 0. \tag{38}$$

On the other hand, the corresponding wave functions can be found by referring to table 1 and the relation (34) as

$$F(s) = s^{\alpha_{12}} e^{\alpha_{13} s} L_n^{\alpha_{10}-1}(\alpha_{11} s)$$
$$= N_{nJ} s^{\frac{1}{\alpha}\sqrt{J(J+1)c_0 + (m^2 - E^2)r_0^2}} e^{-\frac{s}{\alpha}\sqrt{J(J+1)c_2 + D^2 r_0^2}}$$
$$\times L_n^{\left(\frac{2}{\alpha}\sqrt{J(J+1)c_0 + (m^2 - E^2)r_0^2}\right)}\left(\frac{2}{\alpha}\sqrt{J(J+1)c_2 + D^2 r_0^2} \, s\right), \tag{39}$$

where $N_{nJ}$ is the normalization constant. We remember that $s = e^{-\alpha x} = e^{-\alpha(r-r_0)}$ and thus one can find $F(r)$. From (16a), (16b) and (16c), the other spinor components of the wave function can be found as

$$G(r) = \frac{1}{m}(E - U_V^0) F(r), \tag{40a}$$

$$H_1(r) = -\frac{\alpha_J}{m}\left(\frac{d}{dr} - \frac{J+1}{r}\right) F(r), \tag{40b}$$

$$H_{-1}(r) = -\frac{\varsigma_J}{m}\left(\frac{d}{dr} + \frac{J}{r}\right) F(r). \tag{40c}$$

### 4.1. A Few Special Cases

Here we consider a few special cases of much interest. First, when the total angular momentum is being zero, i.e. $J = 0$, Eq. (38) reduces to

$$(2n+1)D\alpha r_0 - 2EDr_0^2 + 2Dr_0^2 \sqrt{m^2 - E^2} = 0 \tag{41}$$

Second, we consider the non-relativistic limiting case when $\varepsilon \approx E - m$ and $2m \approx E + m$ where the non-relativistic energy $\varepsilon$ is such that $|\varepsilon| \ll m$. Therefore, by using transformations as: $J \to l$, $E^2 - m^2 \to 2m\varepsilon$, $(U_V^0)^2 - 2EU_V^0 \to -2mU_V^0$ and $F(r) \rightarrow R(r)$, then Eq. (17) reduces to the radial Schrödinger equation for the exponential potential-like as

$$\left[\frac{d^2}{dr^2} + 2m\varepsilon + 2mDe^{-a(r-r_0)} - \frac{l(l+1)}{r^2}\right]R(r) = 0. \qquad (42)$$

Thus, one can easily obtain the energy equation for the nonrelativistic case as

$$(2n+1)\alpha\sqrt{l(l+1)c_2} + \left(l(l+1)c_1 - 2mDr_0^2\right) + 2\sqrt{l(l+1)c_2\left(l(l+1)c_0 + 2mEr_0^2\right)} = 0. \qquad (43)$$

## 5. Concluding Remarks

In this paper, we approximately solved the Duffin–Kemmer–Petiau equation for a vector exponential-like decaying potential with any $J \neq 0$ case. We used the Pekeris approximation scheme to deal with the total angular momentum centrifugal term. The parametric generalization of the NU method is used to obtain the energy eigenvalues and the corresponding spinor components of the wave functions. Also, we considered the cases of the zero total angular momentum $J = 0$ and the non-relativistic limit of our solution.

**Acknowledgments**

We would like to thank the kind referees for the positive suggestions and critics which have greatly improved the present manuscript.

**Table 1.** The specific values for the parametric constants used in calculating the energy eigenvalues and wave functions

| Constant | Analytical value |
|---|---|
| $\alpha_1$ | 1 |
| $\alpha_2$ | 0 |
| $\alpha_3$ | 0 |
| $\alpha_4$ | 0 |
| $\alpha_5$ | 0 |
| $\alpha_6$ | $\frac{1}{\alpha^2}\left[J(J+1)c_2 + D^2 r_0^2\right]$ |
| $\alpha_7$ | $\frac{1}{\alpha^2}\left[J(J+1)c_1 - 2EDr_0^2\right]$ |
| $\alpha_8$ | $\frac{1}{\alpha^2}\left[J(J+1)c_0 + (m^2 - E^2)r_0^2\right]$ |
| $\alpha_9$ | $\frac{1}{\alpha^2}\left[J(J+1)c_2 + D^2 r_0^2\right]$ |
| $\alpha_{10}$ | $1 + \frac{2}{\alpha}\sqrt{J(J+1)c_0 + (m^2 - E^2)r_0^2}$ |
| $\alpha_{11}$ | $\frac{2}{\alpha}\sqrt{J(J+1)c_2 + D^2 r_0^2}$ |
| $\alpha_{12}$ | $\frac{1}{\alpha}\sqrt{J(J+1)c_0 + (m^2 - E^2)r_0^2}$ |
| $\alpha_{13}$ | $-\frac{1}{\alpha}\sqrt{J(J+1)c_2 + D^2 r_0^2}$ |
| $\xi_1$ | $\frac{1}{\alpha^2}\left[J(J+1)c_2 + D^2 r_0^2\right]$ |
| $\xi_2$ | $\frac{1}{\alpha^2}\left[-J(J+1)c_1 + 2EDr_0^2\right]$ |
| $\xi_3$ | $\frac{1}{\alpha^2}\left[J(J+1)c_0 + (m^2 - E^2)r_0^2\right]$ |